# Sub-Terahertz Nearfields for Electron-Pulse Compression

Yarden Mazor and Ofer Kfir

School of Electrical Engineering, Tel Aviv University, Tel Aviv, 69978, Israel

Corresponding email: kfir@tauex.tau.ac.il

## Abstract

The advent of ultrafast science with pulsed electron beams raised the need in controlling the temporal features of the electron pulses. One promising suggestion is the nano-selective quantum optics with multi-electrons, which scales quadratically with the number of electrons within the coherence time of the quantum system. Terahertz (THz) radiation from optical nonlinear crystals is an attractive methodology to generate the rapidly varying electric fields necessary for electron compression, with an advantage of an inherent temporal locking to laser-triggered electrons, such as in untrafast electron microscopes. Longer (picosecond-) pulses require sub-THz field for their compression, however, the generation of such low frequencies require pumping with energetic optical pulses and their focusability is fundamentally limited by their mm-wavelength. This work proposes electron-pulse compression with sub-THz fields directly in the vicinity of their dipolar origin, thereby avoiding mediation through radiation. We analyze the merits of nearfields for compression of slow electrons particularly in challenging regimes for THz radiation, such as small numerical apertures, micro-joule-level optical pump pulses, and low frequencies. This sheme can be implemented within the tight constraints of electron microscopes and reach fiels of a few kV/cm below 0.1 THz at high repetition rates. Our paradigm offers a realistic approach for controlling electron pulses spatially and temporally in many experiments, opening the path of flexible multi-electron manipulation for analytic and quantum sciences.

# Introduction

Electron microscopes are an indispensable analytic tool, bringing analytic imaging down to an atomic resolution. The rise of the laser-triggered electron (e⁻-)microscope [1–4] added ultrafast dynamics capabilities to the sub-nm spatial sensitivity [5–8], and opened a viable path towards quantum entanglement between free electrons and photons [9–11]. In such instruments, the temporal features of the electron pulses determine the accessible physics. Shrinking the duration of e⁻-pulses can be typically done using RF cavities [12–14] and laser-driven THz fields [15–18]. Alternatively, optical fields can structure the intra-pulse features dramatically by PINEM (Photon-induced nearfield electron microscopy) [19–22]. Under intense laser illumination, the electron forms periodic micro-pulses as short as a femtosecond or even reaching the attosecond scale, separated by the optical-phase cycle [23–26]. Investigation of quantum aspects of PINEM revealed it constitutes a coherent modulation of the e⁻-wavefunction, which energetically extended to exchanges of hundreds of photons [27–29], can shape the e⁻-wavefunction spatially [30–32], and enable subwavelength field holography [33,34]. Such sub-optical-cycle tailoring of a nano-focused e⁻-beam is a novel *probe* that is phase-locked to the dynamics driven by the same laser [35,36].

A recent, potentially transformative, theoretical prediction suggests exerting quantum-optical *control* at the atomic scale if electron pulses are bunched both globally and internally. The FEBERI scheme (Free-electron bound-electron resonant interaction) [37–39] claims that if multiple electrons shaped to trains of attosecond pulses pass by a quantum system, they can induce coherent excitation nonlinearly at a frequency defined by the micro-pulse separation, that is, the cycle of the PINEM-driving laser. For a two-level system, the transition amplitude is predicted to be proportional to the number of FEBERI-structured electrons, N. Thus, the transition probability scales as N² or, more generally, as the $\sin^2(g_{Qu}N)$ of a Rabi-oscillation cycle, where $g_{Qu}$ is the quantum coupling. Doing so within an electron microscope could allow the manipulation of individual quantum systems with high spatial selectivity, in free space and without any physical probe. The bunch duration matters. The N FEBERI electrons should arrive within the coherence time of the quantum system for their contribution to build up coherently. Temporal compression can enable access to drive short-lived excitations and compensate for the e⁻-pulse Coulomb broadening [40,41]. Hence, FEBERI has a particular set of constraints: (i) high-quality beam for nanoscopic focusing (ii) laser-electron interaction for PINEM (iii) electron compression.

Using few- or single-cycle laser-pumped THz pulses for compressing the electrons is appealing for integrating within an ultrafast electron microscope since it is compact, inherently timed with laser-triggered e⁻-pulses, and a THz cycle fits the duration of short electron pulses (~200-700 fs [40–42]). Intense terahertz waves are generated from the optical rectification of short pulses in lithium niobate ($LiNbO_3$). The radiation forms off-axis beams which are collected and re-focused with high-numerical-aperture (high-NA) optics onto the target. The geometry of the pumping laser pulse, the crystal, and the THz collection play an intricate role in optimizing the THz throughput. For a given pump energy and duration the chosen geometry is dictated by the limit on the peak intensity, due to multi-photon absorption. At 1μm pump wavelength, the limiting intensity is 20-100 GW/cm² [43], above which the THz efficiency diminishes [44]. By focusing a pulse with a tilted front into a $LiNbO_3$ [39,40] prism the electric fields reach above MV/cm in the few-THz regime [47,48]. But since the efficiency of tilted-front pumping drops for pulse energies below the millijoule range [49], it operates at low repetition rates of one or a few kHz. More recent schemes propose THz generation from pulses propagating in a $LiNbO_3$ slab, befitting pulses with up to 200 μJ approximately [50,51]. A slab geometry enables either a compensation of the THz-phase jitter [52,53] or an efficient heat dissipation through its surface [50]. Thus, allowing the THz to be pumped by higher average power, that is, with a higher repetition rate.

While the term THz broadly refers to 0.1-100 THz, only sub-THz is relevant for compressing e⁻-pulses with an initial duration of a few hundred femtoseconds [40]. However, delivering radiation in the sub-THz regime is particularly challenging. A detailed quantitative analysis by Tsarev et al. [50], shows that the radiated power efficiency scales cubically with the THz frequency, and the focused power density scales as the fifth power(!) due to the diffraction limit. For e⁻-beam manipulation, the problem is further exacerbated if high-NA optics cannot be used to reach a diffraction-limited focal spot. Since light can be trivially focused to the one or a few micrometers laser beams can provide dramatically higher energy densities. The beat note of such a tightly focused bi-chromatic laser was suggested as a means to compress a portion of the electrons in a bunch [54]. The few-mm region of addressable e⁻-beam in electron microscopes poses a standing issue as a barrier for compression of e⁻-pulses using THz-fields, especially for e⁻-pulses longer than a picosecond.

This work presents a conceptual change for laser-driven sub-THz compression of ps e⁻-pulses: instead of radiation, having a direct interaction between the electrons and the laser-induced dipolar nearfields. Avoiding an intermediary energy conversion to propagating waves omits the unfavorable frequency scaling of generating and transporting sub-THz radiation. We address this topic analytically and numerically. First, the compressive strength of nearfields from µJ-level pulses in $LiNbO_3$ is compared against optimal radiation and refocusing of THz. The analytic comparison is conducted for quasi-static nonlinear polarization induced in $LiNbO_3$ by the infrared driving pulse. The approximation holds for electron energies below 5 keV within the chosen parameter regime but provides a rough estimation up to tens of keV. The calculation is benchmarked for optical pulses with 1 µJ energy and a frequency of 0.1 THz (100 GHz). For lower frequencies and small NAs, the nearfield-based e⁻-compression is better by an order of magnitude due to the favorable frequency scaling. We describe an optical pumping scheme that maintains the process efficiency for more energetic infrared pumping. Second, we show numerical calculations that quantify the e⁻-compression by THz nearfields. As an example, we find that fields in this approach can reach 2.4 kV/cm at the challenging regime of 0.1 THz, pumped above the intensity that would saturate radiative THz. However, we emphasize that our motivation is not to reach the highest THz field, but rather to find a laser-locked approach with a favorable scaling for experiments with tight constraints. We believe that this small-scale scheme opens a path towards in-situ focusing of e⁻-pulses which is imperative for the coherent interaction of multi-electrons with nanoscopic quantum systems.

The outline of this paper is as follows: first, we present the proposed geometry and the analytic derivation for the compressive force using the nonlinear dipolar nearfields in $LiNbO_3$ and compare the analytic results to the full numerical calculation. Then we compare nearfields to radiation-based electron compression and find the regimes for which the nearfields are superior. We finish by suggesting extensions at higher pumping energies which are unique to the nearfields approach.

# Results and discussion

## *Analytic derivation*

THz generation with µJ infrared laser pulses in LiNbO$_3$ is optimal when implemented with a slab geometry, with a silicon output coupler, where the radiation is ideally collected and refocused by high-NA optics. The temporal profile of the e⁻ energy gain and the resulting e⁻-pulse compression is compared between a direct interaction with the nearfield (Figure 1a) and an optimal scenario of radiation from such a slab (Figure 1b), which interacts with the electron pulse on a distant membrane. To eliminate higher-frequency components we consider a 10-ps-long laser pulse focused near the surface of a Y-cut LiNbO$_3$ crystal, where the c-axis parallel to the surface. The slab geometry allows for efficient cooling of the LiNbO$_3$, which handles the thermal load of a high-repetition-rate laser operation and suppresses absorption by thermal phonons [50,55].

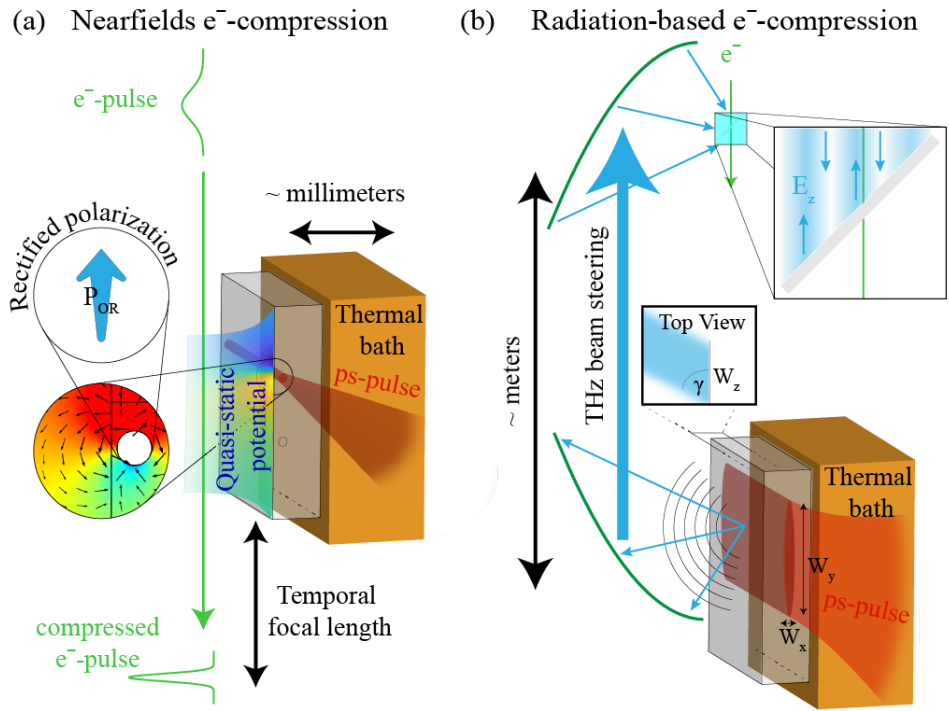

Figure 1 - Illustration of the proposed paradigm for electron compression. (a) The quasi-static nearfields of an optical rectification polarization ($P_{OR}$) in LiNbO3 interact with a traversing electron. The inset shows the calculated electric field (arrows) and potential (colormap) near the induced dipole. The temporally varying field induces a linear velocity-time correlation that compresses the electrons to a short pulse downstream. (b) A typical scheme for e⁻-pulse compression by a THz-irradiated membrane. We consider a slab LiNbO$_3$ crystal pumped by an elliptical optical beam (FWHM semiaxis marked $W_x$, $W_y$), from which a THz field is emitted at an angle γ and transferred with high-NA optics.

The analytic derivation of the e⁻-pulse compression by nearfields vs. THz radiation required several assumptions. (i) The electron propagates along the z-axis of the semi-infinite LiNbO$_3$ crystal. The crystal spans the region $z, y \in [-\infty, \infty]$ and $x \in [-\infty, 0]$. (ii) The fields surrounding the laser-induced polarization are quasi-static. We define the regimes for which this assumption holds at a later point. (iii) The problem is two-dimensional. The confocal parameter of the laser and the distance to its entrance to the crystal are much larger than the spacing between the e⁻-trajectory and the crystal. (iv) The spatial profile of the optical pulse is assumed to be a perfect cylinder, creating a flat-top polarization region with a radius $R_0$ throughout the crystal depth. Thus, we account only for the two-dimensional dipolar density, without considering the laser beam diffraction or possible interference that a Gaussian pulse would have in proximity to the crystal edge. (v) The temporal profile of the optical pulse is a Gaussian with a full-width-at-half-maximum (FWHM) of $T_{FHWM} = 10\ ps$.

Importantly, when approximations are required to keep the calculation analytic, we consistently underestimate the effect of the nearfields while overestimating the radiation. Thus, our quantitative comparative conclusions here are conservative and can be considered the *lowest bound*.

For generating the nonlinear sub-THz polarization, we consider a laser with a peak intensity of $I_0 = 20\frac{GW}{cm^2}$ in the LiNbO$_3$ with a spatial profile of a 25-µm-diameter cylinder and a temporal FWHM span of 10 ps, which has an energy of 1 µJ. Thus, these parameters are a convenient reference. The polarization induced by the optical rectification along the crystal c-axis is $P_{OR}(t) = 2d_{eff}\frac{2I(t)}{cn_p}$, where $n_p$ is the laser's refractive index, $d_{eff}$ is the nonlinear coefficient of LiNbO$_3$ [56,57], $c$ is the speed of light and $I(t) = I_0 \exp(-4\ln 2\ t^2/T_{FWHM}^2)$ is the instantaneous optical intensity. The term $\frac{2I(t)}{cn_p}$ is the square of the electric field that creates the difference-frequency polarization (eq. (1) and eq. (23), ref. [58]) times the vacuum permittivity, $\varepsilon_0$. As illustrated in Figure 1a, the electron propagates purely in the vacuum subspace, $x > 0$. The quasi-static potential is calculated using the image dipole contribution [59], $\Phi(x > 0, z, t) = -\frac{1}{2\pi\varepsilon_0\rho^2}\frac{2}{1+\varepsilon_r}(\vec{p}_{2d}(t) \cdot \vec{\rho})$. $\vec{\rho}$ is the two-dimensional radius vector from a dipole to the point of observation, $\rho^2 = x^2 + z^2$ is the square of the radius vector length, $\varepsilon_r = n_{THz}^2$ is the relative dielectric constant of LiNbO$_3$, $\vec{p}_{2d}(t) = \pi R_0^2 P_{OR}(t)\hat{z}$ is the macroscopic nonlinear polarization per unit depth, measured in Coulombs, and $\hat{z}$ is the unit vector pointing parallel to the crystal's c-axis. The field component relevant for e$^-$-pulse compression is the spatial derivative of the quasi-static potential along the propagation axis,

$$E_z(x,z,t) = -\frac{d\Phi(x,z,t)}{dz} = -\frac{d}{dz}\left(\frac{z}{x^2+z^2}\right)\frac{1}{2\pi\varepsilon_0}\frac{2}{1+\varepsilon_r}\pi R_0^2 \cdot 2d_{eff}\frac{2I(t)}{cn_p}. \quad (1)$$

The electron on-axis acceleration depends on the energy it accumulates throughout its path,

$$U_e(\tau) = (-q)\int_{path} E_z(x,z_{(t)},t)dz_{(t)}. \quad (2)$$

Here, $q$ is the electron charge and $z_{(t)}$ marks the electron trajectory, simplified as one-dimensional. The energy gain varies with the electron timing $\tau$, and its derivative, $dU_e(\tau)/d\tau$ is the figure of merit for e$^-$-pulse compression. We also refer to this figure of merit as the compressive strength since a higher value shortens the resulting e$^-$-pulse duration and the necessary propagation for reaching a full compression. $\tau$ is the relative delay between the electron passage and the optical pulse, such that $\tau = 0$ represents an electron at $z = 0$ when the nearfield is maximal. For an electron traveling with velocity $v_e$ along the z axis, the trajectory is $z_{(t)} = v_e(t-\tau)$. Temporally, the field exists for approximately $T_{FHWM}$, during which the electron passes a finite distance, $v_e T_{FWHM}$. Thus, we simplify the integral by assuming a constant field within the period of $v_e T_{FWHM}$, $E_z(L_d, z, \tau) \to \frac{1}{2}E_z(L_d, 0, \tau)$, resulting in an underestimated energy gain of $U_e(\tau) = (-q)v_e T_{FWHM}\frac{1}{2}E_z(L_d, 0, \tau)$. The maximal value of $dU_e(\tau)/d\tau$ for a temporal Gaussian envelope is $\left.\frac{dU_e}{d\tau}\right|_{max} = e^{-\frac{1}{2}}\sqrt{8\ln 2}\frac{U_{e,max}}{T_{FWHM}} = e^{-\frac{1}{2}}\sqrt{8\ln 2}\, qv_e\frac{1}{2}E_z(L_d,0,0)$. Figure 2a shows the $E_z$ component (log scale) along the electron path for energies up to 40 keV for the analytic approximation. Figure 2b shows the numerically calculated fields. Substituting the optical rectification dipole we find

$$\left.\frac{dU_e^{nearfield}}{d\tau}\right|_{max} = 2v_e e^{-\frac{1}{2}}\sqrt{8\ln 2}\, q\, \frac{1}{1+\varepsilon_r}\left(\frac{R_0}{L_d}\right)^2 \frac{d_{eff}}{\varepsilon_0 n_p c} I_p. \qquad (3)$$

Figure 2c shows the energy accumulated as the electron traverses the laser-driven nearfield region, numerically. It is calculated for an electron with a kinetic energy of 1 keV. Each curve represents a different timing, $\tau$. The final energy, $U_e^{nearfield}(\tau)$, is approximately a Gaussian (see Figure 2d), matching the laser pulse envelope. This stems from the e⁻-energy being well within the regime that matches the quasi-static approximation (see curve coalescence in Figure 2e). The inflection point of the energy-gain shifts temporally from that of the optical pump (See Figure 2d), i.e., the quasistatic approximation, however, we ignore this constant timing shift since it is trivially compensated for by delaying the optical pump. Figure 2e compares the compression figure of merit between the quasistatic approximation and a full dynamical calculation of the fields near the surface of the crystal surface. The parameters are optical pumping with 1 µJ energy and a FWHM duration of 10 ps. We express the results in terms of the optimal (shortest) temporal focal length, $L_{focus}$, which is given by $L_{focus} = \left(\frac{dU_e}{d\tau}\big|_{max}\right)^{-1} \gamma_r^3 \beta^3 c^3 \frac{m_e}{q}$. This focal length is the spatial propagation at which the electron-pulse duration is compressed to a minimum if it was initially dispersionless [60]. Thus, it is a useful parameter in designing experimental layouts. Here, $\beta$ is the unitless relativistic parameter for the velocity, $\gamma_r$ is the relativistic Lorentz factor, and $m_e$ is the electron mass. Figure 2e shows that for kinetic energies below 5 keV the focal length reduces dramatically (note the logarithmic scale) and the exact calculation of the fields in COMSOL converges to the quasi-static calculation. In the following, we use an example of electrons accelerated to 1 keV ($v_e$=0.0625c), well within the approximation's validity range. The exact calculation (red) exhibits a resonant-like deviation from the quasi-static effect (blue) at 11 keV, originating from the match between the electron velocity at these energies and the sub-THz wave velocity in LiNbO$_3$ (0.204c). Overall, the quasi-static approach provides a good estimation for the compression. For electrons up to 30 keV, the quasi-static calculation deviates by a factor of 3 at its worst.

Although the nearfield compressive scheme is non-harmonic, a comparison to radiation fields necessitates a definition of an effective frequency. An ideal compression of an electron pulse requires a uniform compressive strength, that is, a uniform energy-gain gradient. Thus, we define the frequency by the duration for which the compressive strength is higher than 90% of the maximum value. For harmonic fields, 1/7 of the cycle complies with such a requirement. Thus, we calculate the region for which the nearfield's energy gain gradient is $d_\tau U_e > 0.9(d_\tau U_e)_{max}$, and define that duration as 1/7 the effective period. For a Gaussian temporal profile with a FWHM duration $T_{FWHM}$, the criterion is met for $0.26\,T_{FWHM}$, so the effective THz frequency is $f_{THz}^{eff} = 0.55/T_{FWHM}$. Thus, nearfields pumped by 10 ps pulses can compress e⁻-pulses as long as 2.6 ps (see marking on Figure 2d).

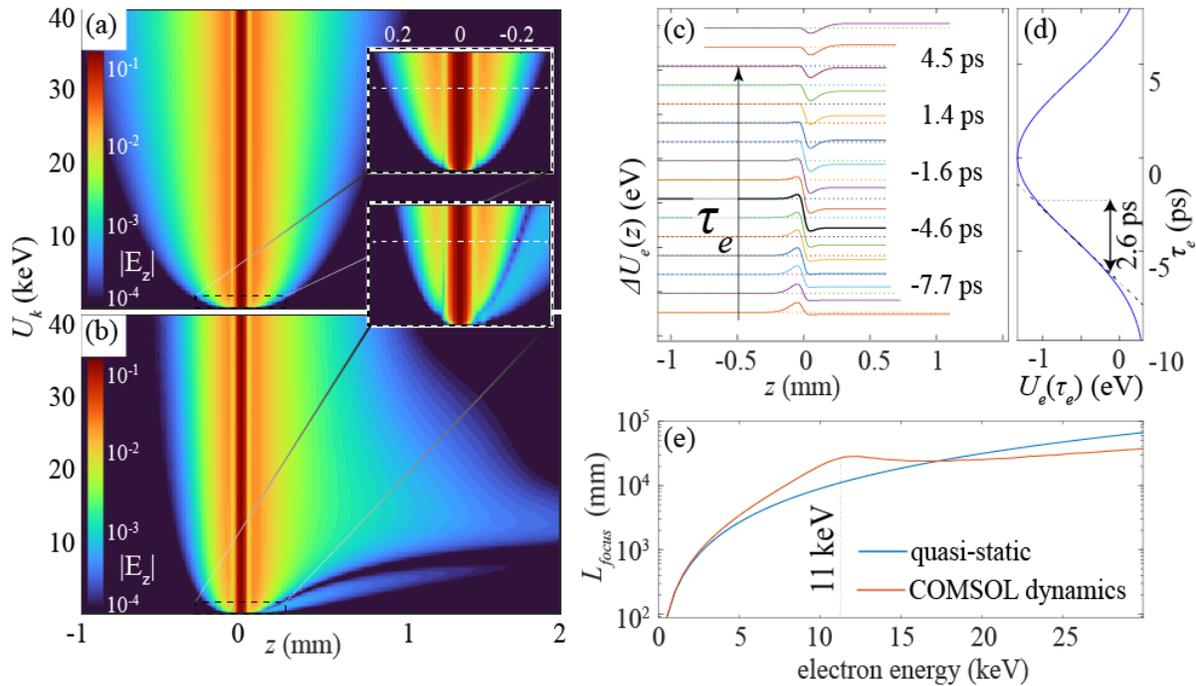

Figure 2 – Compression dynamics of electrons in the nearfield. The graphs describe the reference example, pumped by a peak optical power of 20 GW/cm², pulse duration of 10 ps, focal diameter of 25 µm located in LiNbO3, 50 µm from the electron trajectory. (a) : Quasistatic and (b) full dynamic calculation of the parallel component of the electric field $E_z$ (absolute value, logarithmic color scale) vs. the distance from the dipole and the electron kinetic energy. The dashed white line marks 1 keV on the zoomed insets. The field distribution in (b) is asymmetric due to the radiation field. (c) The cumulative energy gain for electrons traveling through the nearfield vs. their propagation coordinate. Each curve marks different electron timing, $\tau_e$, with respect of the peak energy gain (or loss). (d) The final e⁻-energy gain $U_e$ vs $\tau_e$. The region with $dU_e/d\tau_e > 90\%$ of the maximum spans 2.6 ps. A dashed black line tangent to the maximal slope is added for clarity. Both (a)-(b) describe electrons at a kinetic energy of 1 keV. (e) Temporal focus length ($L_{focus}$, logarithmic scale) for the compression of e⁻-pulse vs. their kinetic energy. The focal length extracted from the time-domain calculation of the electric fields in COMSOL™ (red curve) converges to the quasi-static approximation (blue curve) at low electron energies. The propagation length is as short as a few cm for e⁻-energies <1 keV at 1 µJ optical pump energy. At 11 keV the electron velocity matches the THz wave velocity, 0.204 c, corresponding to a refractive index of 4.9 in the LiNbO3.

## Nearfields vs. radiation for e⁻-pulse compression

To define the relative improvement of the nearfields we find an explicit closed-form expression for the figure of merit for e⁻-compression from THz radiation, based on a slab source at optimal conditions,

$$\left.\frac{dU_e^{rad}}{d\tau}\right|_{max} = 2qv_e \frac{(NA)_{THz} d_{eff} I_p W_x}{\varepsilon_0 \lambda_{THz}^2 c n_p} \sqrt{32\pi^3 T_{rad} \frac{W_z W_y}{n_{THz} \cos\gamma}}. \quad (4)$$

The detailed derivation is in Appendix B. $W_x, W_y, W_z$ are the spatial width of the pump beam and $\gamma$ is the off-axis angle of the THz emission, all of which are marked in Figure 1b. $T_{rad}$ is the idealized transmission coefficient of the THz power from the LiNbO$_3$ crystal to free space, $(NA)_{THz}$ is the numerical aperture of the THz focusing optics, and $\lambda_{THz}$ is the THz wavelength in a vacuum. Thus, the ratio between the approaches for e⁻-pulse compression is

$$\frac{d_\tau U_e^{rad}}{d_\tau U_e^{nearf}} = \frac{(NA)_{THz} W_x}{\lambda_{THz}^2} \sqrt{32\pi^3 T_{rad} \frac{W_z W_y}{n_{THz} \cos\gamma}} \left(e^{-\frac{1}{2}}\sqrt{8\ln 2}\frac{1}{1+\varepsilon_r}\left(\frac{R_0}{L_d}\right)^2\right)^{-1}. \quad (5)$$

We can now consider a specific scenario and acquire the added value of the nearfield approach, quantitatively. For LiNbO$_3$ at a frequency below 0.1 THz $n_{THz} = 4.9, \varepsilon_r = n_{THz}^2$, and $\cos\gamma = 2.3/4.9$. The direct-incident power outcoupling is $T_{rad} = 0.43$, thus, the above ratio is $\frac{d_t U_e^{rad}}{d_t U_e^{nearf}} \approx 238 \frac{(NA)_{THz} W_x \sqrt{W_z W_y}}{\lambda_{THz}^2} \left(\frac{L_d}{R_0}\right)^2$ (see Appendix B). The radiative contribution is governed by the focusing conditions and the source size, while the nearfield effect is governed by one parameter, the distance of the e⁻-path from the crystal surface with respect to the radius of the nearfield dipole, $L_d/R_0$. Note that the electron velocity is implicitly reflected in $L_d$, which is the distance for which the nearfield of a static dipole is approximately constant over a length $v_e T_{FWHM}$. Since a dipole field flips its sign 45° from its maximum, the compressive force scales quadratically for a reduced $L_d$ as long as $L_d \geq v_e T_{FWHM}$. For the evaluation of the radiation's figure of merit, $d_\tau U_{e,ref}^{rad}$ per µJ at 0.1 THz, we estimate the optical pump dimensions. The optical intensity is assumed to peak at 20 GW/cm² since at that level the THz conversion efficiency is quadratic in LiNbO$_3$ (See fig. 5 in Ref. [43]). The 4-photon absorption length for the laser is $L_{4ph} = (\delta_{4ph} I^3)^{-1} = 416$ mm, far longer than a typical crystal[1]. The crystal length is assumed to be 10 mm. Transversely, the laser spot should extend to $W_y > 2\lambda_{THz} = 6$ mm collect the radiation from the resulting 3-mm-wide THz source with a NA<0.5 optics (60° collection angle). The LiNbO$_3$ should be thin with respect to the THz absorption length $\alpha_{THz}$ [56], such that it is weakly affected by propagating at angle $\gamma$ through the slab, hence, $W_x < \frac{\sin\gamma}{\alpha_{THz}} \sim 0.5$ $mm$. For these parameters the optical pulse energy is $E_p = 7200$ $\mu J$ (see Appendix B). Thus, per 1 $\mu J$, the ratio of these cases for a given focusing numerical aperture is

$$\left[\frac{d_t U_{e,ref}^{rad}}{d_t U_{e,ref}^{nearf}}\right]_{per\ 1\mu J,\ 0.1\ THz} = C_r \frac{d_t U_{e,ref}^{rad}}{d_t U_{e,ref}^{nearf}} \cdot \frac{1}{7200} = 0.33(NA)_{THz}. \tag{6}$$

The correction factor introduced here, $C_r$, accounts for the optical pulse duration that results in the target frequency of 0.1 THz. A short pulse generating THz excitation in a slab produces a central frequency of $\sqrt{2\ln 2}/\pi T_{FWHM}$ [50,61]. For 0.1 THz it requires a pulse duration of 3.75 ps. We mentioned above that the nearfield equivalent frequency is $f_{THz}^{eff} = 0.55/T_{FWHM}$. Thus, the same final effective frequency requires an energy ratio equal to the pulse-duration ratio, $C_r = \frac{T_{FWHM}^{nearfield}}{T_{FWHM}^{rad}} = \frac{0.55\pi}{\sqrt{2\ln 2}} = 1.47$.

Even for the ultimate high-NA focusing optics the electron compressive strength from nearfields is 2.5-fold that of radiation. Importantly, the calculation was *systematically biased* in favor of the radiative approach, so the actual enhancement would be greater. The focusability of the astigmatic THz-beam shape and aberrations in high-NA optics can add an order of magnitude. The calculations are also conservative for the near fields. For example, the nearfield effect can be increased by bringing both the optical laser and the e⁻-beam closer to the vacuum-crystal surface, an effect we leave out of the scope of the numerical examples we brought here.

Since our nearfield approach is beneficial for replacing low radiation frequencies, we turn to find the watershed frequency, for which the effect of the two approaches balances. We will refer to the field's effective frequency or wavelength freely, using their free-space dispersion relation $f_{THz} = c\lambda_{THz}^{-1}$. The relative efficiency scales as $\lambda_{THz}^{-5/2}$ since optimally, $W_y \propto \lambda_{THz}$, and eq. (5) divided by the pump energy is proportional to $\left(\lambda_{THz}^{-2}\sqrt{W_y}\right)^{-1}$. Using the reference case calculated for 0.1 THz per μJ in eq. (6), the ratio between the radiative and nearfield methods is $\left(\frac{f_{THz}}{0.1\,THz}\right)^{5/2} 0.33(NA)_{THz}$. Thus, they balance for

$$f_{Watershed} = \frac{0.1\,THz}{(0.33(NA)_{THz})^{2/5}}. \qquad (7)$$

---

[1] We used the conservative 4-photon absorption coefficient from ref. [50], $\delta_{4ph} = 30 \cdot 10^{-7} \frac{cm^5}{GW^3}$, rather than the $10^{-7} \frac{cm^5}{GW^3}$ of Ref. [43].

As an example of a few focusing geometries, for $(NA)_{THz} = 0.5$, 0.1, and 0.009 the nearfield approach surpasses the radiative one for frequencies below 0.2 THz, 0.39 THz, and 1 THz, respectively. In terms of the e⁻-pulse duration for compression, these effective frequencies support >90% of the maximal gradient for 1/7 of their cycle, therefore, the nearfield approach would be preferable for compressing electron pulses that span 750 fs, 400 fs, and 140 fs, respectively.

As a final point of the analytical comparison, we claim that the nearfield approach for compressing e⁻-pulses can scale linearly with the pump energy by two approaches. The first one is to simply pump harder. Although seemingly trivial, radiation sources rely on the macroscopic dipole induced throughout the optical pulse propagation and hence their efficiency suffers from 4-photon absorption for intensities above 20 GW/cm² [43]. However, the nearfield acts on the electron directly and locally, over mere tens of microns, thus, the optical penetration depth is irrelevant as long as it is sufficiently long to approximate an infinite dipolar cylinder. Thus, characteristic decay lengths, $L_{4ph}$, for intensities 100, 200, and 300 GW/cm² comply with the long-source condition, being 3.3 mm, 416 μm, and 123 μm, respectively. These intensities are far from the conservative parameters we use in this paper, however, they can bring the effective nearfields to a few kV/cm at the challenging sub-THz regime. Importantly, they are experimentally realistic based on the literature on recorded saturation and damage intensities, 400 GW/cm² and 1 TW/cm², respectively [44]. Extrapolating from Figure 2b (that is calculated for 20 GW/cm²) the sub-THz field reaches 2.4 kV/cm for an intensity of 300 GW/cm². We comment that the locally generated heat should be extracted to avoid thermal damage, drift, or expansion due to the average power of a high repetition rate laser.

Alternatively, at a given peak intensity, the optical pumping energy can be increased if the beam is expanded parallel to the crystal surface, forming an ellipse. The extended elliptical pump should be sheared spatiotemporally according to the electron velocity, such that the nearfields are effectively phase-matched with it. Let us take the reference calculation (induced polarization cylinder, diameter 25 µm, $T_{FWHM}$=10 ps, energy of 1 µJ). Stretching the optical mode 10-fold to 0.25 mm allows pumping the nearfields with 10 µJ energy at the same efficiency. The temporal shear for 1 keV electrons over the major semi-axis of the ellipse should be 133 ps. Electron pulses in this example would fully compress within 16.5 mm. The geometric constraints for the e⁻-compression arise from the maximum extent of such a stretch since we assume the e⁻-beam experiences a uniform energy-gain gradient. For an e⁻-beam semi-convergence angle of 10 milliradians, the beam's maximal diameter over a distance of 0.25 millimeters is 2.5 µm. The numerical calculations in Figure 3 shows that $d_\tau U_e$ decays sub-exponentially away from the surface, approximated by a characteristic e⁻¹ decay length of 63 µm (red circles). An exponential line is added as a reference. Thus, a uniform interaction can be extended to a few millimeters, allowing the energy efficiency of the nearfield scheme to be maintained up to hundreds of micro joules. The blue crosses in Figure 3 show that the e⁻-pulse duration can be longer if the e⁻-beam passes further away from the LiNbO$_3$. Thus, the compressive force can be traded off for an effective lower frequency, and as mentioned above, for accommodating faster electrons. This spatiotemporal spread of the optical pump and intensities above the 4-photon threshold can be combined, for example, by using a smaller beam closer to the LiNbO$_3$ surface and stretched to improve heat dissipation. New methodologies for ultrafast THz-field mapping by optical microscopy, such as QFIM [62], could quantify experimentally the local sub-THz fields that are presented in Figure 2. Since our calculation in this work is conservative, we expect that such a comparison would reveal that nearfields are better than the above predictions.

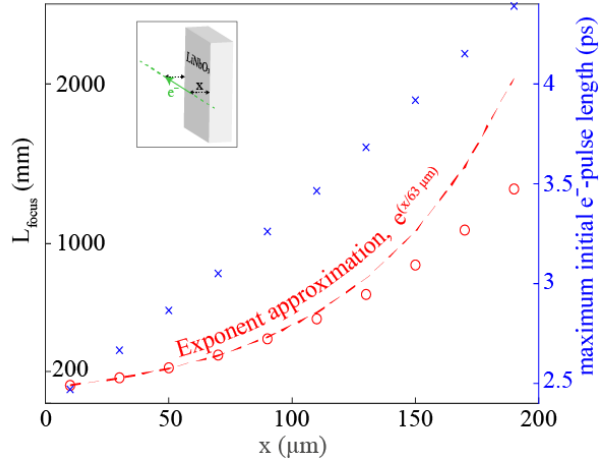

Figure 3 – e⁻-beam compression dependence of the distance from the LiNbO$_3$ surface. When an e⁻-pulse moves parallel to the surface the compressive power decays sub-exponentially with the spacing from the crystal, approximated by an exponent with an e$^{-1}$ length of 63 µm, thus the temporal focal length increases.

## Conclusion

We propose a novel approach for compressing electron pulses using laser-driven fields, exploiting the nearfields emanating from the optically driven crystal directly instead of relying only on re-focused radiated power. Our study shows that analytical quasi-static approximation can be applied for electrons accelerated to below 5 keV (14% the speed of light), assuming an instantaneous dipolar field induced by laser polarization near the surface of a LiNbO$_3$ crystal. The analytical comparison demonstrates that at few-µJ pulse energy, nearfields are especially advantageous for sub-THz frequencies and small numerical apertures. We also present a tilted-pulse method to match the velocity of the electron, which keeps the effectiveness of the nearfields for laser-pump energies of hundreds of µJ. This approach addresses challenges in producing sub-THz fields in confined regions, with inherent laser-locking and elevated saturation intensities. We believe that these effectively intense sub-THz fields would be bridging a gap in controlling electrons, such as compression, deflection, and acceleration. The e⁻-wavefunction manipulation it enables could be the necessary path for exerting nonlinear optics operation with electrons in free space on nano-confined quantum systems.

## Funding

This research was supported by the Israel Science Foundation (grant No. 1021/22, 1089/22)

## Acknowledgment

O.K. gratefully acknowledges the Young Faculty Award from the National Quantum Science and Technology program of the Israeli Planning and Budgeting Committee.

## Disclosures

The authors declare no conflicts of interest.

## Data availability

Data underlying the results presented in this paper are not publicly available at this time but may be obtained from the authors upon reasonable request.

# Appendices

## *Appendix A - Evaluating compression using THz radiation from a LiNbO$_3$ slab*

This section derives this figure of merit for a radiative THz. We do so by calculating the radiated power and the resulting field at a tight focus, followed by integrating over the electron trajectory. We consider a source based on LiNbO$_3$ slab, having a silicon prism on one of its facets for efficient coupling to free-space radiation (see Figure 1b and refs. [50,61]). The beam shape of the driving laser is a transverse ellipse and constant over its propagation. We mark its width, height, and propagation length in the crystal as $W_x$, $W_y$ and $W_z$, respectively, as in ref. [50]. We use the axes notations as the standard in the literature, where $W_z$ is along the laser pulse's propagation direction, whereas in the nearfield calculation the laser propagates along the y-axis. To simplify the quantitative calculation, we assume a rectangular intensity profile, flat within these dimensions. We further assume a collection of the radiation emitted towards the positive x-axis. Note that the emission is tilted towards the z-axis by an angle $\gamma$, determined by phase matching between the THz radiation and the laser-pulse group velocity in the LiNbO$_3$, $\cos\gamma = \frac{n_{p,gr}}{n_{THz}}$ [45]. The angle $\gamma$ is marked on the inset in Figure 1b. For a wavelength of $1030\ nm$ in LiNbO$_3$ the group index is $n_{p,gr} = 2.3$ [63] and we use the measured sub-THz refractive index in LiNbO$_3$, $n_{THZ} = 4.9$ [64], hence $\gamma = 62^o$. We assume optimal handling of the radiation, that is, a perfect out-coupling (e.g., with a silicon prism), aberration-less collimation at high-NA, and an ideal focusing of the radiated THz power into a Gaussian focal spot. The temporal peak of the radiated power can be calculated as a difference-frequency generation (DFG) source (see Table 39 for the DFG efficiency in Ref. [58]),

$$P_0 = \frac{8\pi^2 d_{eff}^2}{\varepsilon_0 n_p^2 n_{THz} c \lambda_{THz}^2} I_p^2 \frac{W_x^2 W_z}{\cos\gamma} W_y.$$

Since the radiation propagates in a tilted angle, the effective medium depth we used for the THz is $W_x/\cos\gamma$, which contributes quadratically. We wrote the above expression in terms of power rather than efficiency, by accounting for the spatial extent of the radiated beam, $W_y \cdot W_z \cos\gamma$ (see inset of Figure 1b). $\lambda_{THz}$ is the THz wavelength in free-space. Our calculation assumes implicitly that the source is effectively two-dimensional and that the THz radiates as a wide beam. In practice sizes larger than a wavelength suffice, as long as they can be collected by optics with a sufficient numerical aperture, that is, $W_z \cos\gamma, W_y > \lambda_{THz}/NA$. Although we assume flat-top pump, in practice, the transverse size of the THz source is $W_y/2$ due to the quadratic nonlinearity. $W_y$ in the above expression is bounded from below by the wavelength, regardless of the NA. For $W_y < \lambda_{THz}$ the source width smaller than $\lambda_{THz}/2$, and hence the radiation geometry is cylindrical and the radiated power scales differently, as $\lambda_{THz}^{-5/2}$. For the out-coupling, we assume an ideal transmission to free space at a direct incidence, yielding a transmission of $T_{rad} = \left(\frac{n_{THz}-1}{n_{THz}+1}\right)^2 = 0.43$.

Once the THz is collimated, further considerations are done to account for the field of the focused beam. The intensity at the waist of a Gaussian beam focused by optics with a numerical aperture $(NA)_{THz}$ is $I_0 = \frac{2\pi (NA)_{THz}^2}{\lambda_{THz}^2} P_0$, hence the peak field is $\sqrt{\frac{4\pi (NA)_{THz}^2}{\varepsilon_0 c \lambda_{THz}^2} P_0}$, or explicitly

$$E_{peak}^{rad} = \frac{(NA)_{THz} d_{eff} I_p W_x}{\varepsilon_0 \lambda_{THz}^2 c n_p} \sqrt{32\pi^3 T_{rad} \frac{W_z W_y}{n_{THz} \cos\gamma}}.$$

As illustrated in Figure 1b, we assume that the perfect THz collection is followed by a perfect refocusing onto a tilted membrane. For simplicity, we account for a 45° tilt, although for electron beams with a finite emittance, there is an optimal angle for transverse velocity matching [17,65]. A traversing electron experiences a sinusoidal acceleration field $E_z^{rad}(z,t) = E_z^{rad}(t) = E_{peak}^{rad} \sin\left(2\pi \frac{t-\tau}{T_{THz}}\right)$. Note that the reflected field is polarized perpendicular to the electron trajectory and does not contribute to an on-axis acceleration. Here $\tau = 0$ represents the electron trajectory that passes the thin surface when the field is nullified. The maximal energy gain is $U_e^{rad} = \int_0^{T/2} q E_{peak}^{rad} \sin\left(\frac{2\pi}{T_{THz}} t\right) v_e dt = 2v_e \frac{T_{THz}}{2\pi} q E_{peak}^{rad}$, and the maximal gradient is $|d_\tau U_e^{rad}| = 2q v_e E_{peak}^{rad}$. Hence we get eq. (4)

$$\left.\frac{dU_e^{rad}}{d\tau}\right|_{max} = 2q v_e \frac{(NA)_{THz} d_{eff} I_p W_x}{\varepsilon_0 \lambda_{THz}^2 c n_p} \sqrt{32\pi^3 T_{rad} \frac{W_z W_y}{n_{THz} \cos\gamma}}.$$

*Appendix B - Quantitative comparison per μJ for radiation and nearfields*

For a two-dimensional Gaussian beam with a peak power $P_0$, the intensity is $I_0 = \frac{2P_0}{\pi \omega_x \omega_y}$, where we consider the FHWM width, which is given by $W_i = \sqrt{2 \ln 2}\, \omega_i$. We can set the intensity to be just below the appearance of 4-photon absorption, $I_0 = I_{4PA} = 20 \frac{GW}{cm^2}$, hence $P_0 = \frac{1}{2}\pi \omega_x \omega_y I_{4PA} = \frac{1}{2}\pi \frac{W_x W_y}{2 \ln 2} I_{4PA}$. Temporally, we assume a Gaussian pulse, where $P_0 = 0.94 \frac{E_p}{T_{FWHM}}$, hence, $E_p = \frac{1}{2}\pi \frac{W_x W_y}{2 \ln 2} I_{4PA} \frac{T_{FWHM}}{0.94}$.

We now substitute $W_x = 0.5\, mm$, $W_y = 6\, mm$, $T_{FWHM} = 10\, ps$ and get $E_p = 0.0072\, J = 7200\, \mu J$.

To combine with eq. (5), and receive a numerical value, we rewrite it,

$$\frac{d_\tau U_e^{rad}}{d_\tau U_e^{nearf}} = \frac{(NA)_{THz} W_x}{\lambda_{THz}^2} \sqrt{32\pi^3 T_{rad} \frac{W_z W_y}{n_{THz} \cos\gamma}} \left( e^{-\frac{1}{2}\sqrt{8\ln 2}} \frac{1}{1+\varepsilon_r} \left(\frac{R_0}{L_d}\right)^2 \right)^{-1}$$

$$= \frac{(NA)_{THz} W_x \sqrt{W_z W_y}}{\lambda_{THz}^2} \left(\frac{L_d}{R_0}\right)^2 \underbrace{\frac{\sqrt{\frac{32\pi^3 T_{rad}}{n_{THz} \cos\gamma}}}{\left(e^{-\frac{1}{2}\sqrt{8\ln 2}} \frac{1}{1+\varepsilon_r}\right)}}_{(*)}.$$

By substituting $n_{THz} = 4.9$, $\varepsilon_r = n_{THz}^2$, $T_{rad} = 0.43$, and $\gamma = 62°$, we get evaluate the term (*) as

$$\frac{\sqrt{\frac{32\pi^3 T_{rad}}{n_{THz} \cos\gamma}}}{\left(e^{-\frac{1}{2}\sqrt{8\ln 2}} \frac{1}{1+\varepsilon_r}\right)} = 238, \text{ therefore,}$$

$$\frac{d_\tau U_e^{rad}}{d_\tau U_e^{nearf}} = 238 \frac{(NA)_{THz} W_x \sqrt{W_z W_y}}{\lambda_{THz}^2} \left(\frac{L_d}{R_0}\right)^2.$$

We can now use the chosen values, $W_z = 10\, mm$, $L_d = 50\, \mu m$, $R_0 = 12.5\, \mu m$, $\lambda_{THz} = 3\, mm$ and get $\frac{W_x \sqrt{W_z W_y}}{\lambda_{THz}^2} \left(\frac{L_d}{R_0}\right)^2 = 6.88$. Thus, the ratio for the energy gain of the radiation vs. nearfield is

$$\frac{d_\tau U_e^{rad}}{d_\tau U_e^{nearf}} = 1642(NA)_{THz}.$$

Thus, the ratio per µJ is 0.228, which when multiplied by $C_r = 1.47$ results in 0.335, as in eq. (6).

# References

[1]  B. Barwick, H. S. Park, O.-H. Kwon, J. S. Baskin, and A. H. Zewail, *4D Imaging of Transient Structures and Morphologies in Ultrafast Electron Microscopy*, Science **322**, 1227 (2008).
[2]  D.-S. Yang, O. F. Mohammed, and A. H. Zewail, *Scanning Ultrafast Electron Microscopy*, PNAS **107**, 14993 (2010).
[3]  S. Meuret et al., *Complementary Cathodoluminescence Lifetime Imaging Configurations in a Scanning Electron Microscope*, Ultramicroscopy **197**, 28 (2019).
[4]  A. Arbouet, G. M. Caruso, and F. Houdellier, *Ultrafast Transmission Electron Microscopy: Historical Development, Instrumentation, and Applications*, in *Advances in Imaging and Electron Physics* (Elsevier Inc., 2018), pp. 1–72.
[5]  T. Danz, T. Domröse, and C. Ropers, *Ultrafast Nanoimaging of the Order Parameter in a Structural Phase Transition*, Science **371**, 371 (2021).
[6]  D. R. Cremons, D. A. Plemmons, and D. J. Flannigan, *Femtosecond Electron Imaging of Defect-Modulated Phonon Dynamics*, Nat Commun **7**, 1 (2016).
[7]  Y. Kurman et al., *Spatiotemporal Imaging of 2D Polariton Wave Packet Dynamics Using Free Electrons*, Science **372**, 1181 (2021).
[8]  M. Solà-Garcia, S. Meuret, T. Coenen, and A. Polman, *Electron-Induced State Conversion in Diamond NV Centers Measured with Pump–Probe Cathodoluminescence Spectroscopy*, ACS Photonics **7**, 232 (2020).
[9]  O. Kfir, *Entanglements of Electrons and Cavity Photons in the Strong-Coupling Regime*, Phys. Rev. Lett. **123**, 103602 (2019).
[10] V. D. Giulio, M. Kociak, and F. J. G. de Abajo, *Probing Quantum Optical Excitations with Fast Electrons*, Optica, OPTICA **6**, 1524 (2019).
[11] A. Feist et al., *Cavity-Mediated Electron-Photon Pairs*, Science **377**, 777 (2022).
[12] T. van Oudheusden, P. L. E. M. Pasmans, S. B. van der Geer, M. J. de Loos, M. J. van der Wiel, and O. J. Luiten, *Compression of Subrelativistic Space-Charge-Dominated Electron Bunches for Single-Shot Femtosecond Electron Diffraction*, Phys. Rev. Lett. **105**, 264801 (2010).
[13] M. R. Otto, L. P. René de Cotret, M. J. Stern, and B. J. Siwick, *Solving the Jitter Problem in Microwave Compressed Ultrafast Electron Diffraction Instruments: Robust Sub-50 Fs Cavity-Laser Phase Stabilization*, Structural Dynamics **4**, 051101 (2017).
[14] G. F. Mancini, B. Mansart, S. Pagano, B. van der Geer, M. de Loos, and F. Carbone, *Design and Implementation of a Flexible Beamline for Fs Electron Diffraction Experiments*, Nuclear Instruments and Methods in Physics Research Section A: Accelerators, Spectrometers, Detectors and Associated Equipment **691**, 113 (2012).
[15] C. Kealhofer, W. Schneider, D. Ehberger, A. Ryabov, F. Krausz, and P. Baum, *All-Optical Control and Metrology of Electron Pulses*, Science **352**, 429 (2016).

[16] D. Zhang et al., *Segmented Terahertz Electron Accelerator and Manipulator (STEAM)*, Nature Photon **12**, 336 (2018).
[17] D. Ehberger, K. J. Mohler, T. Vasileiadis, R. Ernstorfer, L. Waldecker, and P. Baum, *Terahertz Compression of Electron Pulses at a Planar Mirror Membrane*, Phys. Rev. Applied **11**, 024034 (2019).
[18] D. P. Ehberger, Electron Pulse Control with Terahertz Fields, d, MPQ, 2019.
[19] B. Barwick, D. J. Flannigan, and A. H. Zewail, *Photon-Induced near-Field Electron Microscopy*, Nature **462**, 902 (2009).
[20] A. Feist, K. E. Echternkamp, J. Schauss, S. V. Yalunin, S. Schäfer, and C. Ropers, *Quantum Coherent Optical Phase Modulation in an Ultrafast Transmission Electron Microscope*, Nature **521**, 200 (2015).
[21] S. T. Park, M. Lin, and A. H. Zewail, *Photon-Induced near-Field Electron Microscopy (PINEM): Theoretical and Experimental*, New J. Phys. **12**, 123028 (2010).
[22] F. J. García de Abajo, A. Asenjo-Garcia, and M. Kociak, *Multiphoton Absorption and Emission by Interaction of Swift Electrons with Evanescent Light Fields*, Nano Lett. **10**, 1859 (2010).
[23] M. Kozák, N. Schönenberger, and P. Hommelhoff, *Ponderomotive Generation and Detection of Attosecond Free-Electron Pulse Trains*, Phys. Rev. Lett. **120**, 103203 (2018).
[24] Y. Morimoto and P. Baum, *Diffraction and Microscopy with Attosecond Electron Pulse Trains*, Nature Physics **14**, 252 (2018).
[25] K. E. Priebe, C. Rathje, S. V. Yalunin, T. Hohage, A. Feist, S. Schäfer, and C. Ropers, *Attosecond Electron Pulse Trains and Quantum State Reconstruction in Ultrafast Transmission Electron Microscopy*, Nature Photonics **11**, 793 (2017).
[26] A. Ryabov, J. W. Thurner, D. Nabben, M. V. Tsarev, and P. Baum, *Attosecond Metrology in a Continuous-Beam Transmission Electron Microscope*, Science Advances **6**, eabb1393 (2020).
[27] R. Dahan et al., *Resonant Phase-Matching between a Light Wave and a Free-Electron Wavefunction*, Nature Physics **16**, 11 (2020).
[28] O. Kfir, H. Lourenço-Martins, G. Storeck, M. Sivis, T. R. Harvey, T. J. Kippenberg, A. Feist, and C. Ropers, *Controlling Free Electrons with Optical Whispering-Gallery Modes*, Nature **582**, 7810 (2020).
[29] J.-W. Henke et al., *Integrated Photonics Enables Continuous-Beam Electron Phase Modulation*, Nature **600**, 653 (2021).
[30] G. M. Vanacore et al., *Ultrafast Generation and Control of an Electron Vortex Beam via Chiral Plasmonic near Fields*, Nat. Mater. **18**, 573 (2019).
[31] A. Feist, S. V. Yalunin, S. Schäfer, and C. Ropers, *High-Purity Free-Electron Momentum States Prepared by Three-Dimensional Optical Phase Modulation*, Phys. Rev. Research **2**, 043227 (2020).
[32] S. Tsesses, R. Dahan, K. Wang, T. Bucher, K. Cohen, O. Reinhardt, G. Bartal, and I. Kaminer, *Tunable Photon-Induced Spatial Modulation of Free Electrons*, Nat. Mater. **22**, 3 (2023).
[33] I. Madan, G. M. Vanacore, E. Pomarico, G. Berruto, R. J. Lamb, D. McGrouther, T. T. A. Lummen, T. Latychevskaia, F. J. G. de Abajo, and F. Carbone, *Holographic Imaging of Electromagnetic Fields via Electron-Light Quantum Interference*, Science Advances **5**, eaav8358 (2019).

[34] F. Houdellier, G. M. Caruso, S. Weber, M. J. Hÿtch, C. Gatel, and A. Arbouet, *Optimization of Off-Axis Electron Holography Performed with Femtosecond Electron Pulses*, Ultramicroscopy **202**, 26 (2019).

[35] D. Nabben, J. Kuttruff, L. Stolz, A. Ryabov, and P. Baum, *Attosecond Electron Microscopy of Sub-Cycle Optical Dynamics*, Nature **619**, 7968 (2023).

[36] J. H. Gaida, H. Lourenço-Martins, M. Sivis, T. Rittmann, A. Feist, F. J. G. de Abajo, and C. Ropers, *Attosecond Electron Microscopy by Free-Electron Homodyne Detection*, arXiv:2305.03005.

[37] A. Gover and A. Yariv, *Free-Electron–Bound-Electron Resonant Interaction*, Phys. Rev. Lett. **124**, 064801 (2020).

[38] B. Zhang, D. Ran, R. Ianconescu, A. Friedman, J. Scheuer, A. Yariv, and A. Gover, *Quantum Wave-Particle Duality in Free-Electron--Bound-Electron Interaction*, Phys. Rev. Lett. **126**, 244801 (2021).

[39] D. Rätzel, D. Hartley, O. Schwartz, and P. Haslinger, *Controlling Quantum Systems with Modulated Electron Beams*, Phys. Rev. Research **3**, 023247 (2021).

[40] A. Feist et al., *Ultrafast Transmission Electron Microscopy Using a Laser-Driven Field Emitter: Femtosecond Resolution with a High Coherence Electron Beam*, Ultramicroscopy **176**, 63 (2017).

[41] M. Kozák, T. Eckstein, N. Schönenberger, and P. Hommelhoff, *Inelastic Ponderomotive Scattering of Electrons at a High-Intensity Optical Travelling Wave in Vacuum*, Nature Physics **14**, 121 (2018).

[42] K. Wang, R. Dahan, M. Shentcis, Y. Kauffmann, A. Ben Hayun, O. Reinhardt, S. Tsesses, and I. Kaminer, *Coherent Interaction between Free Electrons and a Photonic Cavity*, Nature **582**, 50 (2020).

[43] M. C. Hoffmann, K.-L. Yeh, J. Hebling, and K. A. Nelson, *Efficient Terahertz Generation by Optical Rectification at 1035 Nm*, Opt. Express **15**, 11706 (2007).

[44] A. G. Stepanov, J. Kuhl, I. Z. Kozma, E. Riedle, G. Almási, and J. Hebling, *Scaling up the Energy of THz Pulses Created by Optical Rectification*, Opt. Express, OE **13**, 5762 (2005).

[45] J. Hebling, G. Almási, I. Z. Kozma, and J. Kuhl, *Velocity Matching by Pulse Front Tilting for Large-Area THz-Pulse Generation*, Opt. Express, OE **10**, 1161 (2002).

[46] L. Wang, G. Tóth, J. Hebling, and F. Kärtner, *Tilted-Pulse-Front Schemes for Terahertz Generation*, Laser & Photonics Reviews **14**, 2000021 (2020).

[47] H. Hirori, A. Doi, F. Blanchard, and K. Tanaka, *Single-Cycle Terahertz Pulses with Amplitudes Exceeding 1 MV/Cm Generated by Optical Rectification in LiNbO3*, Appl. Phys. Lett. **98**, 091106 (2011).

[48] J. A. Fülöp, L. Pálfalvi, S. Klingebiel, G. Almási, F. Krausz, S. Karsch, and J. Hebling, *Generation of Sub-MJ Terahertz Pulses by Optical Rectification*, Opt. Lett., OL **37**, 557 (2012).

[49] Q. Tian et al., *Efficient Generation of a High-Field Terahertz Pulse Train in Bulk Lithium Niobate Crystals by Optical Rectification*, Opt. Express, OE **29**, 9624 (2021).

[50] M. V. Tsarev, D. Ehberger, and P. Baum, *High-Average-Power, Intense THz Pulses from a LiNbO3 Slab with Silicon Output Coupler*, Appl. Phys. B **122**, 30 (2016).

[51] S. B. Bodrov, A. N. Stepanov, M. I. Bakunov, B. V. Shishkin, I. E. Ilyakov, and R. A. Akhmedzhanov, *Highly Efficient Optical-to-Terahertz Conversion in a Sandwich Structure with LiNbO$_3$ Core*, Opt. Express, OE **17**, 1871 (2009).

[52] J. Kuttruff, M. V. Tsarev, and P. Baum, *Jitter-Free Terahertz Pulses from LiNbO$_3$*, Opt. Lett., OL **46**, 2944 (2021).
[53] M. I. Bakunov, E. S. Efimenko, S. D. Gorelov, N. A. Abramovsky, and S. B. Bodrov, *Efficient Cherenkov-Type Optical-to-Terahertz Converter with Terahertz Beam Combining*, Opt. Lett., OL **45**, 3533 (2020).
[54] Z. Zhao, K. J. Leedle, D. S. Black, O. Solgaard, R. L. Byer, and S. Fan, *Electron Pulse Compression with Optical Beat Note*, Phys. Rev. Lett. **127**, 164802 (2021).
[55] S.-W. Huang, E. Granados, W. R. Huang, K.-H. Hong, L. E. Zapata, and F. X. Kärtner, *High Conversion Efficiency, High Energy Terahertz Pulses by Optical Rectification in Cryogenically Cooled Lithium Niobate*, Opt. Lett. **38**, 796 (2013).
[56] J. Hebling, A. G. Stepanov, G. Almási, B. Bartal, and J. Kuhl, *Tunable THz Pulse Generation by Optical Rectification of Ultrashort Laser Pulses with Tilted Pulse Fronts*, Applied Physics B: Lasers and Optics **78**, 593 (2004).
[57] R. Sowade, I. Breunig, C. Tulea, and K. Buse, *Nonlinear Coefficient and Temperature Dependence of the Refractive Index of Lithium Niobate Crystals in the Terahertz Regime*, Appl. Phys. B **99**, 63 (2010).
[58] R. L. Sutherland, *Handbook of Nonlinear Optics*, 2nd ed. (CRC Press, Boca Raton, 2003).
[59] H. A. Haus, *Electromagnetic Fields and Energy* (Prentice-Hall, Englewood Cliffs, N.J., 1989).
[60] P. Baum, *Quantum Dynamics of Attosecond Electron Pulse Compression*, Journal of Applied Physics **122**, 223105 (2017).
[61] M. I. Bakunov, S. B. Bodrov, A. V. Maslov, and M. Hangyo, *Theory of Terahertz Generation in a Slab of Electro-Optic Material Using an Ultrashort Laser Pulse Focused to a Line*, Phys. Rev. B **76**, 085346 (2007).
[62] M. B. Heindl, N. Kirkwood, T. Lauster, J. A. Lang, M. Retsch, P. Mulvaney, and G. Herink, *Ultrafast Imaging of Terahertz Electric Waveforms Using Quantum Dots*, Light Sci Appl **11**, 1 (2022).
[63] Polyanskiy Mikhail N., *Refractive Index Database*, http://refractiveindex.info.
[64] M. Unferdorben, Z. Szaller, I. Hajdara, J. Hebling, and L. Pálfalvi, *Measurement of Refractive Index and Absorption Coefficient of Congruent and Stoichiometric Lithium Niobate in the Terahertz Range*, J Infrared Milli Terahz Waves **36**, 1203 (2015).
[65] D. Ehberger, A. Ryabov, and P. Baum, *Tilted Electron Pulses*, Phys. Rev. Lett. **121**, 094801 (2018).